\documentclass[11pt]{amsart}
\usepackage{amsmath}
\usepackage{amssymb}
\usepackage{graphicx}
\newtheorem{theorem}{Theorem}[section]

\newtheorem{lemma}[theorem]{Lemma}

\theoremstyle{definition}

\numberwithin{equation}{section}

\title[On the Near-Critical Behavior of Continuous Polymers]{On the Near-Critical Behavior of Continuous Polymers}

\author[ L. Koralov]{Leonid Koralov}

\address[L. Koralov]{Dept of Mathematics, University of Maryland,
College Park, MD 20742}
\email{{\tt  koralov@math.umd.edu}}

\author[S. Molchanov]{Stanislav Molchanov}
\address[S. Molchanov]{Dept of Mathematics and Statistics, UNCC, NC 28223 and National Research Univ., Higher School of Economics, Russian Federation}
\email{\tt smolchan@uncc.edu}

\author[B. Vainberg]{Boris Vainberg}
\address[B. Vainberg]{Dept of Mathematics and Statistics, UNCC, NC 28223}
\email{\tt brvainbe@uncc.edu}



\begin{document}

\begin{abstract}
The aim of this paper is to investigate the distribution of a
continuous polymer in the presence of an attractive finitely
supported potential. The most intricate behavior can be observed
if we simultaneously and independently vary two parameters: the temperature, which
approaches the critical value, and the length of the polymer chain,
which tends to infinity.
%In earlier work, we identified the
%distributions that appear in the limit (after a diffusive scaling
%of the original polymer measures). We assumed a specific relation
%between the two parameters. In this paper,
We
%allow the parameters to vary independently and
describe how the typical size of the polymer depends on the two
parameters.
\end{abstract}

\maketitle

\section{Introduction}

At the critical value of the temperature, polymers exhibit a transition between folded (globular) and unfolded states. In a variety of chemical and biological publications, the molecule in its folded or nearly-folded state is modeled by a ball, whose radius $r(\beta, t)$
depends on the temperature (we'll denote the inverse temperature by $\beta$)  and on the number of monomers comprising the molecule (we will
refer to this quantity as length and will denote it by $t$), which is assumed to be large. In our model, the distance along the polymer will play the same role as
the time variable in parabolic equations, hence the notation~$t$.

When the polymer molecules are observed in a liquid with sufficiently low Reynolds number,  the radius affects the diffusion coefficient of a
molecule (denoted by $D(\beta,t)$), with the relationship provided by the Stokes-Einstein equation,
\[
D(\beta,t) = \frac{c}{\beta r(\beta,t)}~,
\]
where $c$ is a constant determined by the properties of the media.  This formula allows one to predict the behavior of polymers in a liquid and also provides an approach to study $r(\beta, t)$ experimentally, by measuring
$D(\beta,t)$
% for a protein and for a reference molecule with known $r(\beta,t)$
(\cite{WGS}, \cite{MFN}). The goal of this paper is to
study the diameter $r(\beta,t)$ (and, consequently, $D(\beta, t)$) as a
function two independent vatiables when $\beta$ approaches its critical value and
$t$ tends to infinity.

% When the molecules are observed in a liquid with sufficiently low Reynolds number,  the radius affects the
%diffusion coefficient of a molecule (denoted by $D(\beta,T)$), with the relationship provided by the Stokes-Einstein equation,
%\[
%D(\beta,T) = \frac{c}{\beta r(\beta,T)}~,
%\]
%where $c$ is a constant determined by the properties of the media. The radii of polymers and their diffusion coefficients in different media (in relation to %temperature and the length of the polymer) are studied in many (mostly experimental) biological papers (see, e.g., \cite{MFN} and references therein).
%We examine a mathematical model that allows for
%detailed analysis of the diameter (and, consequently, the diffusion coefficient) as $T$ approaches its critical value, independently of $L$, which is assumed to
%tend to infinity.

We consider the ``mean field" type model (also called the deterministic pinned model)
of long homogeneous polymers (see, e.g., \cite{LGK}), where the polymer chain interacts with the external attractive
potential.
Thus the statistical weight of a realization of a polymer depends
on the proximity to a given point (that plays the role of the expected center of mass of the polymer).
%Namely, assume that we have a nonnegative,
%not identically equal to zero, compactly supported function $v \in C_0^{\infty}(\mathbb{R}^n)$ and a coupling constant $\beta\geq 0$
%(inverse temperature).
Namely, the distribution of the polymer is given by the Gibbs measure  $\mathrm{P}_{\beta,t}$ with a nonnegative, not
identically equal to zero potential  $v \in C_0^{\infty}(\mathbb{R}^d)$  and a
coupling constant (inverse temperature) $\beta \geq 0$, which regulates the strength of the attraction.  The precise definition of the measure will be given below, but here we note that it
is a measure  on  the space
$C([0,t],\mathbb{R}^d)$, whose elements $\omega(\cdot)$ are interpreted
as realizations of the polymer on $[0,t]$. It follows from the Feynman-Kac formula that, under the
measures $\mathrm{P}_{\beta,t}$, the processes $\{\omega(s), 0\leq s\leq
t\}$ are time-inhomogeneous and Markovian and that their
transition densities can be expressed in terms of the fundamental
solution $p_{\beta}$ of the parabolic equation
\begin{equation} \label{frst}
\frac{\partial u}{\partial t}=H_{\beta}u, ~~~{\rm where}~~~~
H_{\beta}=\frac{1}{2}\Delta  +\beta v: L^2(\mathbb{R}^d)\to
L^2(\mathbb{R}^d).
\end{equation}
It was shown that in $d\geq 3$, at a certain positive (critical) value of
the coupling constant, a phase transition occurs
between a densely packed globular state and an extended state of
the polymer. Namely, for a fixed value of $\beta > \beta_{\rm cr}$ (globular state), a typical
polymer realization remains at a distance of order one from the origin
as $t\to\infty$. On the other hand, for $\beta \leq \beta_{\rm
cr}$ (extended state), typical
polymer realizations are at a distance of order $\sqrt{t}$ from the origin, and thus need to be scaled by the factor $\sqrt{t}$
in the spatial variables in order to get a non-trivial limit as $t \rightarrow \infty$. The critical value $\beta_{\rm cr}$ of the coupling
constant
coincides \cite{CKMV09} with the spectral bifurcation point for operator $H_\beta$: the spectrum is absolutely continuous and coincides with semi-axis $(-\infty,0]$ when $\beta<\beta_{\rm cr}$, and additional positive eigenvalues exist when $\beta>\beta_{\rm cr}$, see also \cite{PV}. Thus
\[
\beta_{\rm cr}=\sup\{\beta>0|\sup\sigma(H_{\beta})=0\},
\]
where  $\sigma(H_{\beta})$ is the spectrum of the operator  $ H_{\beta}$.

We'll focus on the case when $d = 3$. The radius of the polymer can be defined
in terms of the second moment $\sigma_{\beta, t}^2$ of the distance of the end-point from the origin,
\[
r(\beta, t)  = \sigma_{\beta, t}.
\]
As we explain below, the latter can be
expressed in terms of the fundamental solution $p_{\beta}=p_{\beta}(t,0,x)$ of problem (\ref{frst}) with initial data $p_{\beta}|_{t=0}=\delta_0(x)$ as follows:
\begin{equation} \label{smom}
\sigma^2_{\beta, t} = \frac{\int_{\mathbb{R}^3} |x|^2 p_{\beta}(t,0,x)dx }{  \int_{\mathbb{R}^3}p_{\beta}(t,0,x)dx}.
\end{equation}

For two functions $f$ and $g$, we'll write $f \approx g$ if
there are positive constants $c$ and $C$ such that $cf \leq g \leq Cf$ for all the values of the variables that are sufficiently close
to their asymptotic values (in our case, $\beta$ is sufficiently close to $\beta_{\rm cr}$  and $t$ is sufficiently large).

The main result of the paper is the following.
\begin{theorem} \label{mtee}
The radius of the polymerhas the following asymptotic
behavior when $|\beta - \beta_{\rm cr}|$ is small and $t$ is large:
\begin{equation} \label{th}
r({\beta, t}) \approx \left\{\begin{array}{c}
                            {(\beta-\beta_{\rm cr})^{-1}} ~~~ if ~~ (\beta-\beta_{\rm cr})\sqrt {t}\geq1,\\
                            ~~~~\sqrt{t}~~~~~~~~~if ~~ (\beta-\beta_{\rm cr})\sqrt {t}\leq1 .
                          \end{array}
													\right.
\end{equation}
\end{theorem}

\noindent
{\bf Remark.} The case when one takes $t \uparrow \infty$ first and then $\beta \downarrow \beta_{\rm cr}$ has been analyzed in the physics literature (see, e.g.,
\cite{GK}, Chapter 1, Section 7). The corresponding relation has the form $r \approx (\beta-\beta_{\rm cr})^{-1}$ and can be viewed as a sub-case of the first part of formula~(\ref{th}). However,  this relation may lead to a wrong result if one forgets that it is valid for the thermodynamic limit (where $t=\infty$). For example, if $\beta-\beta_{\rm cr}\sim 1/t$,  the relation $r \approx (\beta-\beta_{\rm cr})^{-1}$ would give $r\approx t$ instead of $r\approx\sqrt{t}$, which follows 
from (\ref{th}). In fact, our result provides the asymptotics
in two variables simultaneously, without assuming any relationship between $t$ and $\beta$.

Moreover, in applications, it is often important to consider large, but not infinite, values of $t$.

Observe, also, that (\ref{th}) implies that $r({\beta, t}) \approx \sqrt{t}$ as $t \uparrow \infty$ and
$\beta \uparrow \beta_{\rm cr}$. It seems that this fact has not been
addressed in the physics literature.
\\

\noindent
{\bf Remark.} The arguments in the proof allow one to specify the coefficients in the asymptotic formula above. We provide here only the expressions in two limiting cases:
\[
r({\beta, t}) \sim \alpha_+(\beta) (\beta-\beta_{\rm cr})^{-1}  ~~~ {\rm if} ~~(\beta-\beta_{\rm cr})\sqrt {t}\to\infty,
\]
where $\alpha_+(\cdot)$ is smooth on $[\beta_{\rm cr},\beta_{\rm cr}+\varepsilon]$ and $\alpha_+(\beta_{\rm cr}) = \sqrt{3/\varkappa}$, where $\varkappa$ is defined below in (\ref{l0}), and, for each $C$,
\[
 r({\beta, t}) \sim  \alpha_-((\beta-\beta_{\rm cr})\sqrt t)t ~~~ {\rm if} ~~ |\beta-\beta_{\rm cr}|\to 0,~t\to\infty,~(\beta-\beta_{\rm cr})\sqrt {t}\leq C,
\]
where $\alpha_-(\cdot)$ is smooth on $\mathbb{R}$ and $\alpha_-(-\infty) = \sqrt{3}$. The asymptotics above are uniform in $\beta$ in the first case and in $(\beta-\beta_{\rm cr})\sqrt t$ in the second one.
\\

\noindent
{\bf Remark.} In a particular case, when the potential $v$ is the indicator function of a ball centered at the origin, a similar result was proved in \cite{KMV}. There,
the analysis was based on the explicit formula for the fundamental solution, which allows for more or less explicit calculation of the moments.
\\

Let us now give the precise definition of the Gibbs measure and briefly discuss some of the earlier results most relevant to our paper.
Let the space
$C([0,t],\mathbb{R}^3)$  be equipped with the Wiener measure
$\mathrm{P}_{0,t}$. The elements $\omega(.)$ of the space are interpreted
as realizations of a continuous polymer on $[0,t]$ and are
distributed according to the Gibbs measure $\mathrm{P}_{\beta,t}$
with
\[
\frac{d\mathrm{P}_{\beta,t}}{d\mathrm{P}_{0,t}}(\omega)=
\frac{e^{\beta\int_0^tv(\omega(s))ds}}{Z_{\beta,t}},\qquad
\omega\in C([0,t],\mathbb{R}^3),
\]
where
\[
Z_{\beta,t}=\mathrm{E}_{0,t}e^{\beta\int_0^tv(\omega(s))ds}
\]
is the partition function.
%Thus the polymer measure we consider is of a ``mean field" type,
%where the polymer chain interacts with the external attractive
%potential (as in \cite{LGK}).
%While the potential is assumed to be
%constant in time in our paper, many interesting results have been
%obtained for disordered media, i.e., time-dependent random
%potentials (existence of phase transitions, dependence of the
%growth rate of the partition function on the temperature, etc.,
%see \cite{AS}, \cite{Gia}, for example). However,
%the disordered models are, most likely, not amenable to such a detailed analysis
%of the phase transition phenomena as we propose.
It follows from the Feynman-Kac formula that the finite-dimensional distributions of $\mathrm{P}_{\beta,t}$ are
\begin{equation}  \label{fdd}
\begin{split}
& \mathrm{P}_{\beta,t}(\omega(t_1)\in A_1,...,\omega(t_n)\in A_n) \\
& =\frac{1}{Z_{\beta,t}}\int_{A_1}...\int_{A_n}
\int_{\mathbb{R}^3}p_{\beta}(t_1,0,x_1)...p_{\beta}(t-t_n,x_n,y)
dydx_n...dx_1
\end{split}
\end{equation}
for $0\leq t_1\leq...\leq t_n\leq t$ and $A_1,...,A_n\in \mathcal{B}(\mathbb{R}^3)$, where $p_{\beta}$ is the fundamental
solution  of equation (\ref{frst}). It is worth noting that
\[
Z_{\beta,t}=\int_{\mathbb{R}^3}p_{\beta}(t,0,y)dy.
\]
The expression (\ref{smom}) for the second moment now follows from the definition of the measure~$\mathrm{P}_{\beta,t}$.

In \cite{CKMV09}, we used the detailed analysis of the spectral structure of partial differential operators with a compactly supported potential to
describe the distribution of long polymer chains for each fixed value of $\beta$, including $\beta_{\rm cr}$. Subsequently, our results were
generalized and adapted to several related models: the case of power-law decay of the potential at infinity (Lacoin \cite{La}), the case of the
underlying operator being the generator of a stable process (Takeda, Wada \cite{TW}, Li, Li \cite{LiLi}, Nishimori \cite{Ni}), the case of zero-range potentials (our own work \cite{CKMV10}, \cite{KP}, Fitzsimmons, Li \cite{FL}), etc.

Let us contrast the result of the current paper with the results of the closely related work \cite{KP}.
In \cite{KP}, we considered the situation when $\beta = \beta(t)$ is such that
\begin{equation} \label{rela}
(\beta(t)-\beta_{\rm cr})\sqrt{t} \rightarrow \chi \in \mathbb{R}~~~{\rm as}~~{t \rightarrow \infty}.
\end{equation}
 It was shown
that, after scaling the time by $t$ and the spatial variables by $\sqrt{t}$,
the measures $\mathrm{P}_{\beta(t),t}$ converge, as $t \rightarrow \infty$, to certain limiting measures.
The limiting measures~$ \mathrm{Q}_{\chi}$ were introduced in
\cite{CKMV10} as the polymer measures  on $C([0,1],\mathbb{R}^3)$
corresponding to zero-range attracting potentials (i.e., the
potentials that are, roughly speaking, concentrated at the
origin). In the current paper, we do not make the assumption (\ref{rela}), and the two parameters $\beta$ and $t$ can vary independently.
The scaling required to get a nontrivial limit now depends on the relationship between the parameters.

\section{Proof of the main result}
We represent $p_\beta$ in the form
\begin{equation} \label{pu}
p_\beta=p_0+u, \quad {\rm where} \quad
p_0=\frac{e^{-|x|^2/2t}}{(2\pi t)^{3/2}}
\end{equation}
and $u$ is the solution of the problem
\begin{equation} \label{frst1}
\frac{\partial u}{\partial t}=H_{\beta}u+\beta vp_0, ~~~u(0,x)=0.
\end{equation}
Denote by $r_{\lambda,0}$ the Laplace transform of $p_0$, i.e.,
\[
r_{\lambda,0}=\frac{e^{-\sqrt{2\lambda}|x|}}{2\pi|x|}, \quad  \lambda\notin(-\infty,0].
\]
Applying the Laplace transform to (\ref{frst1}), we get
\begin{equation} \label{lap}
u(t,x)=\frac{1}{2\pi i}\int_{a-i\infty}^{a+i\infty}e^{t\lambda}\hat{u}(\lambda,x)d\lambda,
\end{equation}
where $a$ is a positive constant and $\hat{u}\in L^2(\mathbb{R}^3)$ is the solution of the equation
\begin{equation} \label{stati}
(\frac{1}{2}\Delta  +\beta v-
\lambda)\hat{u}=-\beta vr_{\lambda,0}, \quad  \lambda\notin(-\infty,0].
\end{equation}
One can choose any constant $a$ in (\ref{lap}) that is larger than the supremum of the spectrum $\sigma(H_\beta)$ of operator $H_\beta$. It was shown in \cite{CKMV09} that there is a constant $\beta_{{\rm cr}}>0$ such that $\sigma(H_\beta)=(-\infty,0]$ when $\beta<\beta_{{\rm cr}}$ and $\sigma(H_\beta)=(-\infty,0]\bigcup \lambda_0$ when $\beta-\beta_{{\rm cr}}>0$ is small enough, where $\lambda_0=\lambda_0(\beta)$ is a simple positive eigenvalue of $H_\beta$ and $\lambda_0(\beta)\to 0$ as $\beta\downarrow\beta_{{\rm cr}}$. Moreover,
\begin{equation} \label{l0}
\lambda_0(\beta)\sim \varkappa(\beta-\beta_{{\rm cr}})^2 ~~~{\rm as} ~~~ \beta\downarrow\beta_{{\rm cr}}, \quad \varkappa>0.
\end{equation}
Thus we can take $a=1$ when $|\beta-\beta_{{\rm cr}}|$ is small enough.

Let us discuss the convergence of the integral (\ref{lap}). Using the resolvent identity, the solution
$\hat{u}\in L^2(\mathbb{R}^3)$ of (\ref{stati}) can be written in the form
\begin{equation} \label{res}
\hat{u}= - R_{\lambda,\beta} (\beta vr_{\lambda,0})  =- R_{\lambda,0}(I+\beta vR_{\lambda,0})^{-1}(\beta vr_{\lambda,0}),
\end{equation}
where $R_{\lambda,\beta}=(H_\beta-\lambda)^{-1}:L_2(\mathbb{R}^3)\to L_2(\mathbb{R}^3)$,
and $R_{\lambda,0}$
 is the operator of convolution with the function $-r_{\lambda,0}$.
Obviously, $\|r_{\lambda,0}\|_{L_2}\leq C|\lambda|^{-1/2}$ as $|\lambda|\to\infty$, $|{\rm arg}(\lambda)| \leq 3\pi/4$. Since $\|R_{\lambda,\beta}\|_{L_2}$ does not exceed the inverse distance from the spectrum of operator $H_\beta$, it follows that
\begin{equation} \label{inft}
\|\hat{u}\|_{L_2}\leq C|\lambda|^{-3/2} \quad {\rm as}~~ |\lambda|\to\infty,~~ |{\rm arg}(\lambda)| \leq 3\pi/4.
\end{equation}
This justifies the convergence of the integral in (\ref{lap}). Moreover, function $r_{\lambda,0}$ (as an element of $L_2(\mathbb{R}^3)$) is analytic in $\lambda\in \mathbb C'=\mathbb C\backslash(-\infty,0]$, and the operator-function $R_{\lambda,\beta}$ is meromorphic in $\lambda \in \mathbb C'$ with  the only
pole at $\lambda=\lambda_0(\beta)$. Together with (\ref{inft}), this allows us to replace the contour of integration in (\ref{lap}) by the contour $\Gamma(a)$ that is obtained by splitting the contour in (\ref{lap}) into two halves (where Im$\lambda\gtrless0$) and rotating each of them around the point $\lambda=a$ to the left (in the direction of the half-plane Im$\lambda<0$) by angle $\pi/4$. Thus
\begin{equation} \label{intu}
\begin{split}
u(t,x)=\frac{1}{2\pi i} & \int_{\Gamma(a)}  \int_{\mathbb{R}^3}\frac{e^{-\sqrt{2\lambda}|x-y|+\lambda t}}{2\pi|x-y|}f(\lambda,\beta,y)dyd\lambda,  \\
& \quad f:=(I+\beta vR_{\lambda,0})^{-1}(\beta vr_{\lambda,0}).
\end{split}
\end{equation}
We rewrite the latter expression as
\begin{equation} \label{ff}
 f:=(\beta^{-1}+ vR_{\lambda,0})^{-1}( vr_{\lambda,0}).
\end{equation}

Denote by $L_{2,b}(\mathbb{R}^3)$ the space of $L_2$ functions with supports in the ball $B_b=\{|x|\leq b\}$. We fix an arbitrary $b$ such that $B_b$ contains the support of the potential $v$. Obviously, $vr_{\lambda,0}$ is an entire $L_{2,b}(\mathbb{R}^3)$ function of $k=\sqrt\lambda$ and the operator
$vR_{\lambda,0}$ considered in $L_{2,b}(\mathbb{R}^3)$ is an entire operator-valued function of $k$, and the operator is compact for each $k$. From (\ref{ff}) and the analytic Fredholm theorem, it follows that $f(k^2,\beta,\cdot)\in L_{2,b}(\mathbb{R}^3)$ admits a meromorphic continuation from the half plane Re$k>0$ (that corresponds to $\lambda\in \mathbb C'$) to the whole complex $k-$plane, see \cite{V75} for similar statements in more general settings.

Since the integral kernel of $-vR_{k^2,0}$ is nonnegative for all $k\in \mathbb{R}$, the Perron-Frobenius theorem implies that the
operator $-vR_{k^2,0}$ in $L_{2,b}(\mathbb{R}^3)$,
for each real $k\in \mathbb{R}$, has the principal eigenvalue $\sigma_0(k)$ that is simple, positive, with a non-negative eigenfunction $\varphi(k,x)$ and $\sigma_0(k)$ is larger than the real parts of other eigenvalues. Obviously, $\sigma_0(k)$ is an analytic function of $k\in \mathbb{R}$ since the operator $vR_{\lambda,0}$ considered in $L_{2,b}(\mathbb{R}^3)$ has this property and the eigenvalue is simple.

\begin{lemma} \label{l00}The following statements are valid:

1) Operator $\beta^{-1}+ vR_{\lambda,0}$ is invertible in $L_{2,b}(\mathbb{R}^3)$ for non-real $k$ in the half-plane Re$k>0$ when $\beta^{-1}$ is real, and it is invertible for those real $k$ for which $\sigma_0(k)<\beta^{-1}$.

2) $\sigma_0(0)= {1}/{\beta_{\rm cr}}$

3) The function $\sqrt{\lambda_0(\beta)},~ 0<\beta-\beta_{\rm cr}\ll 1$, can be extended analytically to a small neighborhood of the point $\beta=\beta_{\rm cr}$. The
relations
\begin{equation} \label{eq}
\beta^{-1}=\sigma_0(k) \quad  {\it and} \quad  k=\sqrt{\lambda_0(\beta)}
\end{equation}
are equivalent when $|k|$ and $|\beta-\beta_{\rm cr}|$ are small. The function
\begin{equation} \label{var}
\varsigma(k,\beta)=\frac{{\beta}^{-1}-\sigma_0(k)}{k-\sqrt{\lambda_0(\beta)}},
\end{equation}
after extension by continuity at zeroes of the denominator, is analytic in both arguments in a neighborhood of the point $k=0,~\beta=\beta_{\rm cr}.$
Moreover, $c=\varsigma(0,\beta_{\rm cr}) > 0$.
\end{lemma}

\noindent
{\bf Remark.} When $\beta>\beta_{\rm cr}$, relations (\ref{eq}) define the pole $\lambda=k^2$ of the resolvent $R_{\lambda,\beta}$ (i.e., the eigenvalue of $H_\beta$). The poles of the resolvent at negative $k$, given by  (\ref{eq}) when $\beta<\beta_{\rm cr}$, define resonances located at points $\lambda=k^2$ on the second sheet of the Riemann surface of the spectral parameter $\lambda$. Solutions of the corresponding elliptic equation with these values of $\lambda$ grow exponentially at infinity. The pole at $\lambda=k=0$ that appears when $\beta=\beta_{\rm cr}$ corresponds to the ground state of the operator $H_{0,\beta_{\rm cr}}$.
\proof  Since the kernel of the operator $R_{\lambda,0}$ is trivial, the resolvent identity
\begin{equation} \label{resid}
R_{\lambda,\beta} = R_{\lambda,0}(I+\beta vR_{\lambda,0})^{-1}
\end{equation}
implies that each pole (with respect to $k$) of $(\beta^{-1}+ vR_{\lambda,0})^{-1}$ in the half plane Re$k>0$ leads to the pole of the resolvent $R_{\lambda,\beta}$ at $\lambda=k^2\in \mathbb{C}'$. Since the spectrum of $H_\beta$ for real $\beta$ is real, $k$ must be real. Absence of the real poles when $\sigma_0(k)<\beta^{-1}$ is a trivial consequence of the definition of $\sigma_0(k)$. The first statement is proved.

Let us prove the second statement. Note that $\sigma_0(k)$ is a strictly decreasing analytic (and therefore continuous) function of $k\in \mathbb{R}~$  since the integral kernel of the operator $-vR_{k^2,0}$ has this property. Assume that $\sigma_0(0)>\frac{1}{\beta_{\rm cr}}$. Then there is $k'>0$ such that $\sigma':=\sigma_0(k')>\frac{1}{\beta_{\rm cr}}$ and the operator $(\sigma'+ vR_{\lambda,0})^{-1}$ has a pole at $ k=k'$, and therefore, $(I+ \beta'vR_{\lambda,0})^{-1}$ with $\beta'=\frac{1}{\sigma'} < \beta_{\rm cr}$ has a
pole at $ k=k'>0$. Since the kernel of the operator $R_{\lambda,0}$ is trivial, (\ref{resid}) implies that the resolvent $R_{\lambda,\beta'}$ has a pole on the positive semiaxis, and this contradicts the definition of $\beta_{\rm cr}$ as the smallest value of $\beta$ for which $H_\beta$ has a positive eigenvalue. Thus $\sigma_0(0)\leq\frac{1}{\beta_{\rm cr}}$. Assume now that $\sigma_0(0)<\frac{1}{\beta_{\rm cr}}$. Then there exists $\beta'>\beta_{\rm cr}$ such that $\sigma_0(0)<\frac{1}{\beta'}$,
and therefore the operator $-vR_{k^2,0}-\frac{1}{\beta'}$ is invertible for all $k>0$, i.e., $H_{\beta'}$ does not have positive eigenvalues. This again contradicts the definition of $\beta_{\rm cr}$. Thus $\sigma_0(0)=\frac{1}{\beta_{\rm cr}}$.

In order to prove the last statement, we note that (\ref{resid}) and the definition of $\sigma_0(k)$ imply that the resolvent $R_{\lambda,\beta}$ with $\beta>\beta_{\rm cr}$ has a pole at $k>0$, which is related to $\beta$ by the first equation (\ref{eq}), and $\lambda = k^2$ is the eigenvalue of $H_\beta$. The same eigenvalue is given by the second equation when  $\beta-\beta_{\rm cr}$ is positive and small, i.e., these equations are equivalent when $0<\beta-\beta_{\rm cr}\ll 1$ ($k$ is positive in this case). The derivative $\sigma_0'(0)$ can be evaluated using the second relation in (\ref{eq}):
\[
\sigma_0'(0)=-\lim_{\beta\downarrow\beta_{\rm cr}} \frac{1}{ \beta^2 \frac{d}{d \beta} \sqrt{\lambda_0(\beta))} } ~.
\]
 From here and (\ref{l0}) it follows that $\sigma_0'(0) <  0$, i.e., equation ${\beta}^{-1}=\sigma_0(k)$ can be solved for $k$ when $|k|,~|\beta-\beta_{\rm cr}|\ll 1$. The solution $k=k(\beta)$ is analytic in $\beta$ and coincides with $\sqrt{\lambda_0(\beta))}$ when $\beta>\beta_{\rm cr}$. This allows us to extend $\sqrt{\lambda_0(\beta))}$ analytically to a small complex neighborhood of $\beta_{\rm cr}$ and provides the equivalency relations (\ref{eq}) when $|\beta-\beta_{\rm cr}|$ is small.

It remains to note that functions ${\beta}^{-1}-\sigma_0(k)$ and $k-\sqrt{\lambda_0(\beta)}$ are analytic in the pair of variables when $|k|+|\beta-\beta_{\rm cr}|\leq \varepsilon\ll 1$. They have the same set of zeroes and the gradients of these functions are not vanishing. Thus, the ratio of these functions is analytic. The L'Hopital rule implies that $c=\varsigma(0,\beta_{\rm cr})=-\sigma_0'(0) > 0$. The proof is complete. The validity of the remark follows from (\ref{resid}).
\qed

\begin{lemma}\label{lemma1}
If $|\beta-\beta_{\rm cr}|\leq \varepsilon$ and $\varepsilon$ is small enough, then
\begin{equation} \label{le1}
f=\frac{h(k,\beta,x)}{k-\sqrt{\lambda_0(\beta)}},
\end{equation}
where $h$ is an analytic function of $k, \beta$ with values in $L_{2,b}(\mathbb{R}^3)$ when ${\rm Re} k>0$ or $|k|\ll 1$,  which has the following properties:
\[
\|h(\sqrt\lambda,\beta,\cdot)\|_{L_2}\leq C~~ when~~|{\rm arg}(\lambda)| \leq 3\pi/4 ~~ and~~|\beta-\beta_{\rm cr}|~~is~~ {\it sufficiently}~~ small,
\]
and
\begin{equation} \label{inth}
\int_{\mathbb{R}^3}h(0,\beta_{\rm cr},x)dx> 0.
\end{equation}
\end{lemma}
\noindent

\proof We fix a small $\delta>0$. From Lemma \ref{l00} it follows that there is $\varepsilon=\varepsilon(\delta)$ such that the operator $I+\beta vR_{\lambda,0}$ is invertible when Re$k>0, ~|k|\geq\delta$, and $|\beta-\beta_{\rm cr}|\leq \varepsilon$, and therefore the inverse operator $(I+\beta vR_{\lambda,0})^{-1}$ is analytic in $\lambda, \beta$.  Thus from  (\ref{intu}) it follows that $f$ is analytic when Re$k>0, ~|k|\geq\delta,~|\beta-\beta_{\rm cr}|\leq \varepsilon$. Obviously, it can be written in the form (\ref{le1}), where $h=(k-\sqrt{\lambda_0(\beta)})f$ is analytic at the same region.
Since $\|r_{\lambda,0}\|_{L_2}\leq C|\lambda|^{-1/2}$ as $|\lambda|\to\infty$, $|{\rm arg}(\lambda)| \leq 3\pi/4$, and $\|R_{\lambda,\beta}\|_{L_2}$ does not exceed the inverse distance from the spectrum of operator $H_\beta$, from (\ref{intu}) it follows (compare with (\ref{inft})) that
\begin{equation}
\|f\|_{L_2}\leq C|\lambda|^{-1/2} \quad {\rm as}~~ |\lambda|\to\infty,~~ |{\rm arg}(\lambda)| \leq 3\pi/4.
\end{equation}
It remains to justify the analyticity of $h$ when $|k|\leq\delta$ and the validity of (\ref{inth}).

We will use representation (\ref{ff}) when $|k|\leq \delta$. From the definition of $\sigma_0(k)$ for real $k$ and its analyticity in $k$,  it follows that, for $k$ in a sufficiently small circle $|k|\leq\delta$, the operator $vR_{\lambda,0}$  has the simple eigenvalue $\sigma_0(k)$, which depends on $k$ analytically. Thus, from the Laurent expansion in the parameter $\beta^{-1}$, we obtain
\[
(\beta^{-1}+ vR_{\lambda,0})^{-1}=\frac{G(k,\beta)}{\beta^{-1}-\sigma_0(k)},~~~|\beta-\beta_{\rm cr}| \ll 1,
\]
where the operator $G$ is analytic in $k, \beta$ when $|k|\leq\delta,~|\beta-\beta_{\rm cr}|\leq\varepsilon(\delta)$. This formula, together with (\ref{ff}) and
 (\ref{var}), implies (\ref{le1}), where $h$ is analytic when $|k|\leq \delta,~|\beta-\beta_{\rm cr}|\leq\varepsilon(\delta)$, and
 \[
 h(0,\beta_{\rm cr},x)=cG(0,{\beta_{\rm cr}})(vr_{0,0}), \quad c=\varsigma(0,\beta_{\rm cr}) > 0.
 \]
The operator $G(0,{\beta_{\rm cr}})$ is the projection on the ground state $\varphi$ of the operator $vR_{0,0}$. Since $\varphi\geq 0$ and $\varphi$ is positive where $v>0$, and function $vr_{0,0}$ has the same properties, the above formula for $h$ immediately implies (\ref{inth}).
\qed
\\

The rest of the proof of the theorem will be based on formulas (\ref{intu}), (\ref{le1}). Since $f(\lambda,\beta,x)=0$ when $|x|>b$, formula (\ref{intu}) can be rewritten in the form
\[
u(t,x)=\frac{1}{(2\pi)^2i}\int_{\Gamma(a)}\int_{|y|<b}\frac{e^{-\sqrt{2\lambda}|x-y|+\lambda t}}{|x-y|(k-\sqrt{\lambda_0(\beta)})}h(k,\beta,y)dyd\lambda, \quad  k=\sqrt \lambda,
\]
if $|\beta-\beta_{\rm cr}|\leq \varepsilon$ and $\varepsilon$ is small enough.

For any function $w=w(x)$, denote by $w^{(\nu)}$ its moment of order $\nu$, i.e., $w^{(\nu)}=\int_{\mathbb{R}^3}|x|^\nu wdx$. In fact, we will use only $\nu=0$ and $2$. Introducing also (for brevity) the notation $\gamma=\sqrt{\lambda_0(\beta)}$, we obtain that
\begin{equation} \label{intuA}
\begin{split}
u^{(\nu)}=\frac{1}{(2\pi)^2i}\int_{\mathbb{R}^3}\int_{\Gamma(a)} & \int_{|y|<b}\frac{|x|^\nu e^{-\sqrt{2\lambda}|x-y|+\lambda t}}{|x-y|(k-\gamma)}h(k,\beta,y)dyd\lambda dx,  \\
& \quad  k=\sqrt \lambda, \quad  |\beta-\beta_{\rm cr}|\leq \varepsilon.
\end{split}
\end{equation}

Consider first the case $(\beta-\beta_{\rm cr})\sqrt{t}\geq 1$. From this inequality and (\ref{l0}), it follows that $\gamma^2t\geq \varkappa/2$ when
$\beta - \beta_{\rm cr}$ is small (which we ensure by taking a sufficiently small $\varepsilon$).  The integrand in (\ref{intuA}) is meromorphic in $\lambda$ between the contours $\Gamma(a)$ and $\Gamma(\frac{\varkappa}{4t})$ with the only pole at $\lambda=\gamma^2$ (this is the point where $k=\gamma$). The pole is located between the contours since $\gamma^2t\geq\varkappa/2$. The integrand decays exponentially in $\lambda$ at infinity. Hence the contour $\Gamma(a)$ can be moved to $\Gamma(\frac{\varkappa}{4t})$ if we take into account the contribution  from the pole, which will be denoted by $u^{(\nu)}_1$. Thus $u^{(\nu)}=u^{(\nu)}_1+u^{(\nu)}_2$, where $u^{(\nu)}_2$ is given by (\ref{intuA}) with $\Gamma(a)$ replaced by  $\Gamma(\frac{\varkappa}{4t})$, and (after
multiplying the numerator and denominator of the integrand by $k + \gamma$)
\begin{equation} \label{intu1}
\begin{split}
u^{(\nu)}_1=e^{\gamma^2t}\gamma\int_{\mathbb{R}^3}\int_{|y|<b}\frac{|x|^\nu e^{-\sqrt2\gamma|x-y| }}{\pi|x- y |}h(\gamma,\beta,y)dy dx\\
=\frac{e^{\gamma^2 t}}{\gamma^{1+\nu}}\int_{\mathbb{R}^3}\int_{|y|<b}\frac{|z|^\nu e^{-\sqrt2|z-\gamma y|}}{\pi|z-\gamma y|}h(\gamma,\beta,y)dydz \\
=\frac{e^{\gamma^2 t}}{\gamma^{1+\nu}}(C_1+O(\gamma)) \quad {\rm as} \quad \gamma\to 0,
\end{split}
\end{equation}
where
\[
C_1=\frac{1}{\pi}\int_{\mathbb{R}^3}|z|^{\nu-1}e^{-\sqrt2|z|}dz\int_{|y|<b}h(0,\beta_{\rm cr},y)dy
\]
\[
=\frac{(\nu+1)!}{2^{-1+\nu/2}}\int_{|y|<b}h(0,\beta_{\rm cr},y)dy> 0.
\]
We used here the substitution $\gamma x=z$, the uniform (in $\beta$ and $\gamma  = \gamma(\beta)$)
convergence of the resulting double integral, and relation (\ref{inth}). We choose $\varepsilon$ in (\ref{intuA})
so small that $C_1+O(\gamma)\geq \frac{C_1}{2} > 0$ in (\ref{intu1}) when $0<\beta-\beta_{\rm cr}\leq \varepsilon$.

We apply the substitution $(\lambda,x)\to(\mu/t, z\sqrt t)$ in the integral defining $u^{(\nu)}_2$. This leads to
\begin{equation} \label{intu0}
u^{(\nu)}_2=\frac{ t^{\nu/2}}{(2\pi)^2i\gamma}\int_{\mathbb{R}^3}\int_{\Gamma(\frac{\varkappa}{4})}\int_{|y|<b}\frac{|z|^\nu e^{-\sqrt{2\mu}|z-\frac{y}{\sqrt t}|+\mu }}{|z- \frac{y}{\sqrt t}|(\frac{\sqrt\mu}{\gamma\sqrt t}-1)}h(\sqrt{\mu/t},\beta,y)dyd\mu dz.
\end{equation}

We rewrite the coefficient before the integral in the form $\frac{(\gamma^2 t)^{\nu/2}}{\gamma^{1+\nu}(2\pi)^2i}$ so that to make it easier to compare it
with the one in (\ref{intu1}). The integrand and $u^{(\nu)}_2$ can be viewed as functions of independent variables $\xi=1/(\gamma \sqrt{t})$, $\tau=1/\sqrt{t}$, and $\beta$, even
though they can be expressed in terms of $\beta$ and $t$ (or $\gamma$ and $t$).
 Due to Lemma~\ref{lemma1}, the triple integral above converges uniformly on the compact
defined by $0\leq\xi\leq\sqrt{2/\varkappa}$, $0\leq\tau \leq 1$, $0\leq\beta-\beta_{\rm cr}\leq \varepsilon$.  The same is true if the integrand is differentiated in $ \tau$ or $\beta$.  Thus, by using the linear approximation of $u^{(\nu)}_2$ in the variables $\tau$ and $\beta$ (which is uniform in $\xi$),
		 we obtain
\begin{equation} \label{intu2}
\begin{split}
u^{(\nu)}_2 & =\frac{(\gamma^2 t)^{\nu/2}}{\gamma^{1+\nu}}(C_2+O(\frac{1}{\sqrt t}+\gamma)),  \\  C_2 & = \int_{\mathbb{R}^3}\int_{\Gamma(\frac{\varkappa}{4})}\frac{|z|^\nu e^{-\sqrt{2\mu}|z|+\mu }d\mu dz}{(2\pi)^2i|z|(\frac{\sqrt\mu}{\gamma\sqrt t}-1)}\int_{|y|<b}h(0,\beta_{\rm cr},y)dy
\end{split}
\end{equation}
for large $t$ and sufficiently small $\gamma$ (recall that $\gamma=\sqrt{\lambda_0(\beta)}=O(\beta-\beta_{\rm cr})$ as $\beta-\beta_{\rm cr}\to 0$).

Let us show the existence  of positive constants $C_2^-,C_2^+$ such that
\begin{equation} \label{c2}
0<C_2^- \leq C_2 \leq C_2^+ \quad {\rm when} \quad  \gamma^2t\geq \varkappa/2, ~~\nu=0,2.
\end{equation}
Indeed, the integral in $z$ in (\ref{intu2}) can be easily evaluated and,  since $\nu=0,2$ is even, this implies
\begin{align}\nonumber
C_2=C_2(\gamma \sqrt{t},\nu)=\int_{\Gamma(\frac{\varkappa}{4})}\frac{-(\nu+1)! e^{\mu }d\mu }{\pi i(-\sqrt{2\mu})^{2+\nu}(\frac{\sqrt\mu}{\gamma\sqrt t}-1)}\int_{|y|<b}h(0,\beta_{\rm cr},y)dy\\=
\frac{(\nu+1)!c}{2^{1+\nu/2}\pi}\int_{|y|<b}h(0,\beta_{\rm cr},y)dy, \quad {\rm where} \quad  c=\int_{\Gamma(\frac{\varkappa}{4})}\frac{i e^{\mu }d\mu }{\mu^{1+\nu/2}(\frac{\sqrt\mu}{\gamma\sqrt t}-1)}. \label{c3}
\end{align}

The inequalities (\ref{c2}) will follow from (\ref{c3}) and (\ref{inth}) if we show existence of constants $c^\pm$ such that $0<c^-\leq c\leq c^+$. The quantity $c$ is a continuous function of $\gamma^2t$ when $\gamma^2t\geq\varkappa/2$ with a finite limit $c_\nu> 0$ as $\gamma^2t\to\infty$ (the value of $c_\nu$
can be easily evaluated using the residue at $\mu = 0$).
 Hence it remains to show that $c$ is positive for all the values of $\gamma^2t\geq\varkappa/2$.

Let $\nu=0$. Then, splitting the integral into two terms and evaluating the second one using the residue at $\mu=0$, we obtain
\[
c=\int_{\Gamma(\frac{\varkappa}{4}) }\frac{i e^{\mu }(\frac{\sqrt\mu}{\gamma\sqrt t}+1)d\mu }{\mu(\frac{\mu}{\gamma^2 t}-1)}=\int_{\Gamma(\frac{\varkappa}{4}) }\frac{i e^{\mu }d\mu }{\frac{\sqrt\mu}{\gamma\sqrt t}(\frac{\mu}{\gamma^2 t}-1)}+2\pi.
\]
The contour in the integral in the right-hand side can be replaced  by a path going around the cut along the negative semi-axis in the $\mu$ plane, and this leads to the positivity of the first term. Hence, $c >0$ if $\nu=0$. In order to study the case $\nu=2$, we introduce $c(\sigma)$ defined by (\ref{c3}) with the exponent $e^\mu$ replaced by $e^{\mu\sigma}$. By repeating the arguments used in the case of $\nu=0$, we get that $c'(\sigma)>0$ for all $\sigma>0.$
When $\sigma = 0$, the integral can be evaluated through the residue at $\mu = \gamma^2 t$ since the contour can be moved to positive
infinity along the real axis. This leads to $c(0) >0$. Thus, integration in $\sigma$ implies that $c(\sigma)>0$ when $\sigma>0$. Hence $c=c(1)>0$ when $\nu=2$. This completes the proof of (\ref{c2}). Now we choose $\varepsilon$ in (\ref{intuA}) so
small that $C_2+O(\frac{1}{\sqrt t}+\gamma) \geq \frac{C_2}{2}  > 0$ in (\ref{intu2}) when $0<\beta-\beta_{\rm cr}\leq \varepsilon$ and $t$ is large enough.

From (\ref{pu}) it follows that
\[
p_\beta^{(\nu)} =  p_0^{(\nu)} + u^{(\nu)}_1 +  u^{(\nu)}_2.
\]
Since $p_0^{(0)}=1,~p_0^{(2)}=3t$, from (\ref{intu1}), (\ref{intu2}), and (\ref{c2}) it follows that
the last term estimates the other two, and thus
$p_\beta^{(\nu)}\approx\frac{e^{\gamma^2 t}}{\gamma^{1+\nu}}$ when $\frac{1}{t}+|\gamma|\ll 1$ and $\gamma^2 t\geq \frac{\varkappa}{2}$.
This immediately implies (\ref{th}) in the case of $(\beta-\beta_{\rm cr})\sqrt{ t}\geq1$.

Consider now the case when $|\beta-\beta_{\rm cr}|\sqrt{ t}\leq1$. After change of the variables $(\lambda,x)\to(\mu/t, z\sqrt t)$ in (\ref{intuA}), we obtain
\begin{equation} \label{intuC}
u^{(\nu)}=\frac{ t^{(1+\nu)/2}}{(2\pi)^2i}\int_{\mathbb{R}^3}\int_{\Gamma(4\varkappa)}\int_{|y|<b}\frac{|z|^\nu e^{-\sqrt{2\mu}|z-\frac{y}{\sqrt t}|+\mu }}{|z- \frac{y}{\sqrt t}|(\sqrt\mu-\gamma\sqrt t)}h(\sqrt{\mu/t},\beta,y)dyd\mu dz,
\end{equation}
where $\varkappa$ is defined in (\ref{l0})  and $|\beta - \beta_{\rm cr}| \leq \varepsilon$ with a small enough $\varepsilon$. In fact, the contour $\Gamma(at)$ appears in (\ref{intuC}) after the change of the variables. However,  the inequality $|\beta-\beta_{\rm cr}|\sqrt{ t}\leq1$ and (\ref{l0}) imply that $\gamma^2 t\leq 2\varkappa$ when $|\beta-\beta_{\rm cr}|\leq \varepsilon$ and $\varepsilon$ is small enough, i.e.,  the pole $\mu=\gamma^2 t$ of the integrand in (\ref{intuC}) is located to the left of $\Gamma(4\varkappa)$. Hence the integrand is analytic in $\mu$ between $\Gamma(at)$ and $\Gamma(4\varkappa)$. Since it also decays exponentially at infinity, we can use $\Gamma(4\varkappa)$ in (\ref{intuC}) instead of $\Gamma(at)$.

Let us consider the integrand in (\ref{intuC}) and the function $u^{(\nu)}$ as functions of independent variables $\xi=\gamma \sqrt{t}$, $\tau=1/\sqrt{t}$, and $\beta$, in spite of the fact that these variables can be expressed in terms of $\beta$ and $t$ (or $\gamma$ and $t$).
 Then the triple integral converges uniformly (see Lemma~\ref{lemma1}) on the compact
$0\leq\xi\leq\sqrt{2\varkappa}$, $0\leq\tau \leq 1$, $|\beta-\beta_{\rm cr}| \leq \varepsilon$, and therefore, it is continuous there.
On the other hand, $u$ is the solution of the parabolic problem (\ref{frst1}) with the non-negative source $\beta vp_0$, and therefore it is positive for all $t>0$. Hence there are positive constants $C^-,~ C^+$ such that
\[
C^- t^{1+\nu/2}\leq u^{(\nu)}\leq C^+t^{1+\nu/2} ~~~ {\rm when}~~~ \frac{1}{t}\leq 1, ~ |\beta-\beta_{\rm cr}|\leq \varepsilon,~|\beta-\beta_{\rm cr}|\sqrt t\leq 1.
\]
This, (\ref{pu}), and the relation $p_0^{(0)}=1,~p_0^{(2)}=3t$ imply that
 $p_\beta^{(\nu)}\approx t^{(1+\nu)/2}$ when ${t}\to \infty$ and $|\beta-\beta_{\rm cr}|\sqrt t\leq1$. This justifies the statement of the theorem when $|\beta-\beta_{\rm cr}|\sqrt t\leq1$.

 The same argument can be applied to prove the theorem in the case  $(\beta-\beta_{\rm cr})\sqrt t\leq-1$. One needs only to take $|\gamma|\sqrt t$ out of the
integral in (\ref{intuC}):
 \[
u^{(\nu)}=\frac{ t^{(1+\nu)/2}}{(2\pi)^2i|\gamma|\sqrt t}\int_{\mathbb{R}^3}\int_{\Gamma(4\varkappa)}\int_{|y|<b}\frac{|z|^\nu e^{-\sqrt{2\mu}|z-\frac{y}{\sqrt t}|+\mu }}{|z- \frac{y}{\sqrt t}|(\frac{\sqrt\mu}{|\gamma|\sqrt t}+1)}h(\sqrt{\mu/t},\beta,y)dyd\mu dz,
\]
and use the continuity of the integral with respect to
 $\frac{1}{|\gamma|\sqrt t},~\frac{1}{\sqrt t}$, and $\beta$.
\qed
\\
\\\\
\noindent {\bf \large Acknowledgments}:
The work of  L. Koralov was supported by the ARO grant W911NF1710419, the Simons Foundation Fellowship (award number 678928),
and by the Russian Science
 Foundation,  project ${\rm N}^o$~20-11-20119. The work of S. Molchanov was supported by the NSF grant DMS-1714402 and by the Russian Science
 Foundation,  project ${\rm N}^o$~20-11-20119.  The work of B. Vainberg was supported by the NSF grant DMS-1714402 and the Simons
 Foundation grant 527180.
\\
\\

\end{document}